\documentclass[letterpaper,twocolumn,english,prl,superscriptaddress,showpacs,longbibliography,fixfloat,notitlepage]{revtex4-1}
\usepackage[latin9]{inputenc}
\usepackage{amsmath}
\usepackage{amssymb}
\usepackage{graphicx}

\makeatletter

\pdfpageheight\paperheight
\pdfpagewidth\paperwidth

\usepackage[colorlinks,citecolor=darkblue,linkcolor=darkred,urlcolor=darkblue] {hyperref}
\usepackage{xcolor}
\definecolor{darkblue}{rgb}{0.1,0.2,0.6} 
\definecolor{darkred}{rgb}{0.8,0.1,0.2}
\renewcommand{\BibitemShut}[1]{}

\makeatother

\usepackage{babel}
\begin{document}
\global\long\def\E{\mathrm{e}}
\global\long\def\D{\mathrm{d}}
\global\long\def\I{\mathrm{i}}
\global\long\def\mat#1{\mathsf{#1}}
\global\long\def\vec#1{\mathsf{#1}}
\global\long\def\cf{\textit{cf.}}
\global\long\def\ie{\textit{i.e.}}
\global\long\def\eg{\textit{e.g.}}
\global\long\def\vs{\textit{vs.}}
 \global\long\def\ket#1{\left|#1\right\rangle }

\global\long\def\etal{\textit{et al.}}
\global\long\def\tr{\text{Tr}\,}
 \global\long\def\im{\text{Im}\,}
 \global\long\def\re{\text{Re}\,}
 \global\long\def\bra#1{\left\langle #1\right|}
 \global\long\def\braket#1#2{\left.\left\langle #1\right|#2\right\rangle }
 \global\long\def\obracket#1#2#3{\left\langle #1\right|#2\left|#3\right\rangle }
 \global\long\def\proj#1#2{\left.\left.\left|#1\right\rangle \right\langle #2\right|}

\title{(Quasi)Periodic revivals in periodically driven interacting quantum
systems}

\author{David J. Luitz}

\affiliation{Department of Physics, T42, Technische Universität München, James-Franck-Straße
1, D-85748 Garching, Germany}
\email{david.luitz@tum.de}

\author{Achilleas Lazarides}

\affiliation{Max-Planck-Institut für Physik komplexer Systeme, 01187 Dresden,
Germany}
\email{acl@pks.mpg.de}

\author{Yevgeny Bar Lev}

\affiliation{Max-Planck-Institut für Physik komplexer Systeme, 01187 Dresden,
Germany}

\affiliation{Department of Condensed Matter Physics, Weizmann Institute of Science,
Rehovot 76100, Israel}

\affiliation{Department of Chemistry, Columbia University, 3000 Broadway, New
York, New York 10027, USA}
\email{yevgeny.barlev@weizmann.ac.il}

\begin{abstract}
Recently it has been shown that interparticle interactions \emph{generically}
destroy dynamical localization in periodically driven systems, resulting
in diffusive transport and heating. In this work we rigorously construct
a family of interacting driven systems which are dynamically localized
and effectively decoupled from the external driving potential. We
show that these systems exhibit tunable periodic or quasiperiodic
revivals of the many-body wavefunction and thus \emph{of all} physical
observables. By numerically examining spinless fermions on a one dimensional
lattice we show that the analytically obtained revivals of such systems
remain stable for finite systems with open boundary conditions while
having a finite lifetime in the presence of static spatial disorder.
We find this lifetime to be inversely proportional to the disorder
strength.
\end{abstract}
\maketitle
\emph{Introduction.\textemdash{}} Dynamical phases of matter, far
from thermodynamic equilibrium, have recently attracted significant
interest and activity, with periodically driven (Floquet) quantum
many-body systems emerging as one of the main research directions.
Such systems do not thermalize in the conventional sense due to the
continuous injection of energy by the external driving. Nevertheless,
they approach a nonequilibrium steady state (NESS) in which the von
Neumann entropy is maximized, subject to constraints given by the
conservation laws of the system \cite{DAlessio2014,Lazarides2014b,Ponte2014a}.
For generic Floquet systems with no conservation laws, the NESS is
featureless and cannot be locally differentiated from an infinite
temperature state. Some systems can however avoid this fate, due to
the existence of an \emph{extensive} number of conserved quantities
analogous to the Generalized Gibbs Ensemble. Noninteracting systems
\cite{Russomanno2012,Lazarides2014a} and interacting systems with
sufficiently strong disorder \cite{Basko2006a,Lazarides2014,Abanin2014}
are two examples. The nontrivial NESS of these systems potentially
hosts exotic nonequilibrium phenomena such as the recently proposed
time-domain crystalline order \cite{Khemani2015a,Else:2016ue,Else2016b,Ho2017},
which spontaneously breaks the discrete time translation symmetry.
Physical observables in discrete time crystals display subharmonic
oscillations, namely a periodic time dependence with periods longer
than the period of the external drive. Another class of systems which
fail to heat up are noninteracting systems exhibiting \emph{dynamical
localization} (DL) \textemdash{} a complete suppression of transport
\emph{due} to the presence of a special drive \cite{Revie1986}. Generic
interactions destroy DL, leading to diffusive transport and eventually
a featureless high-entropy steady state \cite{Luitz2017b}. However
localization can still be restored by the addition of sufficiently
strong static disorder \cite{Bairey2017}.

In this work we construct a family of periodic \emph{interacting}
systems exhibiting dynamical localization and for which an initial
wave function may show tunable periodic or quasiperiodic revivals.
While the revivals are unstable to generic perturbations of our driving
protocol, we show that in the presence of disorder of strength $W$
they decay on a time-scale proportional to $1/W$. This offers the
possibility of experimental realizations in driven ultracold atoms
in optical lattices. 

\begin{figure}
\includegraphics[width=1\columnwidth]{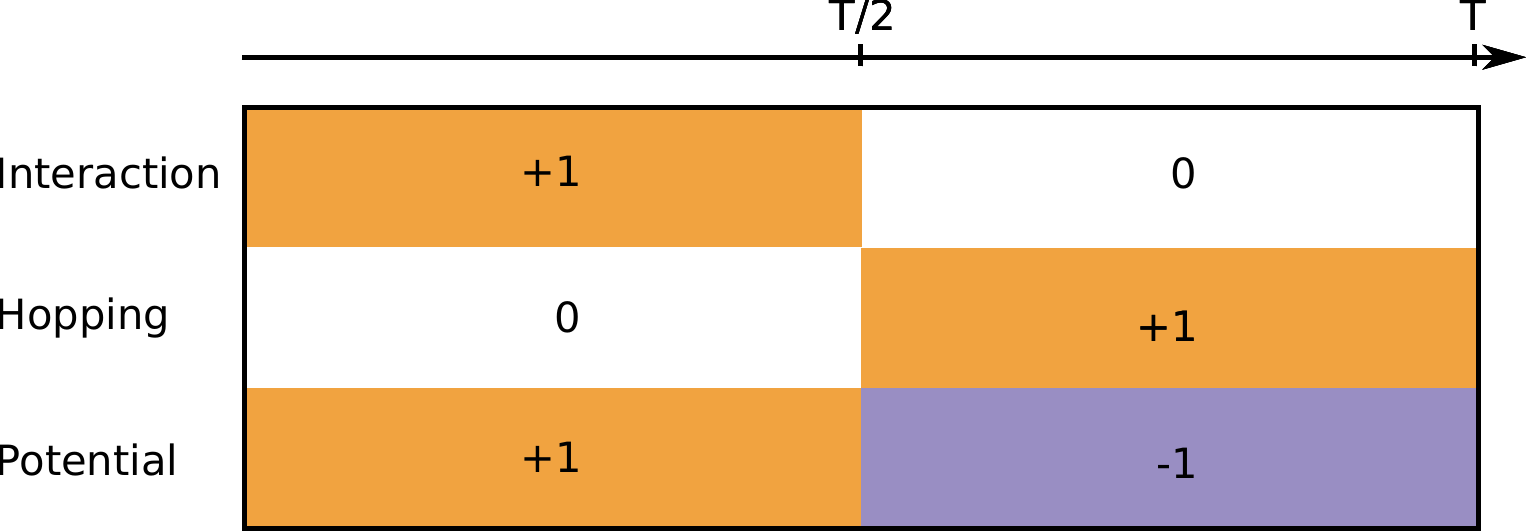}

\caption{\label{fig:Driving-protocol}Driving protocol: The driving is alternated
between a hopping term operating over the first half period, $t\in\left[0,\frac{T}{2}\right)$,
and an interaction term with no hopping operating in the second half
period, $t\in\left[\frac{T}{2},T\right)$. There is also an additional
external periodic potential which for concreteness was chosen to have
an alternating sign over the two half-periods. The figure illustrates
the prefactor and the sign of the different terms.}
\end{figure}

\begin{figure}
\includegraphics{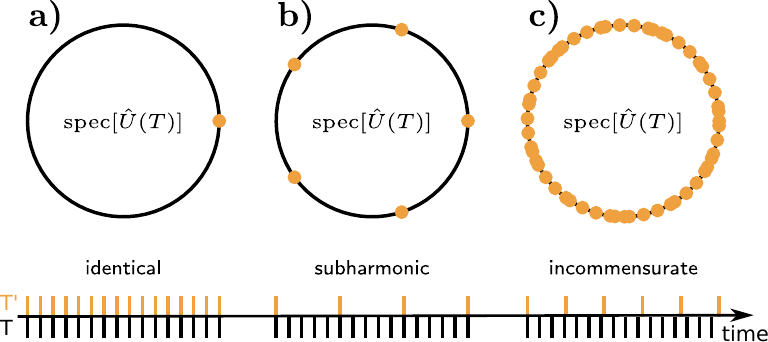}\caption{Spectrum of the Floquet operator $\hat{U}(T)$ on the complex unit
circle. \textbf{a)} Completely degenerate spectrum, $\hat{U}(T)=1$,
with trivial stroboscopic dynamics \textbf{b) }Partially degenerate
spectrum which corresponds to complex unit roots (5th root here).
The system exhibits subharmonic revival, $\hat{U}(5T)=1$. \textbf{c)
}A spectrum which corresponds to a generic quasiperiodic dynamics.
\label{fig:spec}}
\end{figure}

\emph{Theory.\textemdash }We consider a time-dependent $T$-periodic
Hamiltonian $\hat{H}(t+T)=\hat{H}(t)$ where the hopping and the interaction
do not operate at the same time (see Fig.~\ref{fig:Driving-protocol}).
The Hamiltonian is defined over one period as 
\begin{equation}
\hat{H}\left(t\right)=f\left(t\right)\hat{H}_{\text{pot}}+\begin{cases}
\sum_{n,m=1}^{L}h_{n-m}\hat{c}_{n}^{\dagger}\hat{c}_{m} & t\in\left[0,\frac{T}{2}\right)\\
\hat{H}_{\text{int}} & t\in\left[\frac{T}{2},T\right)
\end{cases}.\label{eq:ham}
\end{equation}
Here $L$ is the length of the lattice, $\hat{n}_{m}=\hat{c}_{m}^{\dagger}\hat{c}_{m}$
is the number operator, and $\hat{c}_{m}^{\dagger}$ creates a spinless
fermion at site $m$. The function $f\left(t\right)$ is an arbitrary
periodic function of time with a period of $T=2\pi/\omega$ and an
angular frequency $\omega$. $\hat{H}_{\text{pot}}$ is an external
potential with a constant gradient,
\begin{equation}
\hat{H}_{\text{pot}}=\sum_{m=1}^{L}m\hat{n}_{m}.
\end{equation}
We will further assume that $\hat{H}_{\text{int}}$ commutes with
$\hat{H}_{\text{pot}}$, but is \emph{not} bilinear in $\hat{c}_{m}^{\dagger}$
and $\hat{c}_{m}$, such that the system is interacting. For notational
simplicity we focus here on spinless fermions in one-dimension, though
the same formalism with the appropriate modifications applies also
for bosons and to higher dimensions. It is convenient to work directly
with the propagator of the system,
\begin{equation}
\hat{U}'\left(t,t'\right)\equiv\mathcal{T}\exp\left[-\I\int_{t'}^{t}\mathrm{d\bar{t}}\hat{H}\left(\bar{t}\right)\right],\label{eq:nontranformed_U}
\end{equation}
where $\mathcal{T}$ is the time-ordering operator. Applying the time
dependent unitary transformation
\begin{equation}
\hat{V}\left(t\right)=\E^{\I\mathcal{A}\left(t\right)\hat{H}_{\text{pot}}},\quad\mathrm{with}\mathcal{\quad A}\left(t\right)=\int_{0}^{t}\mathrm{d}\bar{t}\,f\left(\bar{t}\right),\label{eq:unitary_trans}
\end{equation}
 to the propagator yields 
\begin{equation}
\hat{U}\left(t,t'\right)=\hat{V}\left(t\right)\hat{U}'\left(t,t'\right)\hat{V}^{\dagger}\left(t'\right).
\end{equation}
and therefore,
\begin{align}
\I\partial_{t}\hat{U}\left(t,t'\right) & =\left(\I\partial_{t}\hat{V}\left(t\right)\right)\hat{U}'\left(t,t'\right)\hat{V}^{\dagger}\left(t'\right)\\
 & +\hat{V}\left(t\right)\left(\I\partial_{t}\hat{U}'\left(t,t'\right)\right)\hat{V}^{\dagger}\left(t'\right).\nonumber 
\end{align}
Using Eqs.~(\ref{eq:nontranformed_U}) and (\ref{eq:unitary_trans})
we arrive at
\begin{equation}
\I\partial_{t}\hat{U}\left(t,t'\right)=\hat{H}'\left(t\right)\hat{U}\left(t,t'\right),
\end{equation}
where
\begin{align}
\hat{H}'\left(t\right) & \equiv\hat{V}\left(t\right)\hat{H}\left(t\right)\hat{V}^{\dagger}\left(t'\right)-f\left(t\right)\hat{H}_{\text{pot}}.
\end{align}
Since
\begin{equation}
\hat{U}\left(T,0\right)=\mathcal{T}\exp\left[-\I\int_{0}^{T}\mathrm{d\bar{t}}\hat{H}\left('\bar{t}\right)\right].
\end{equation}
Using the assumption $\left[\hat{H}_{\text{int}},\hat{H}_{\text{pot}}\right]=0$,
\emph{the one-period} propagator can be written as $\hat{U}\left(T,0\right)=\hat{U}_{\text{int}}\hat{U}_{\text{nonint}}$,
where 

\begin{equation}
\hat{U}_{\text{nonint}}\equiv\exp\left[-\I\int_{0}^{T/2}\mathrm{d}\bar{t}\left(\sum_{n,m}\E^{\I\mathcal{A}\left(\bar{t}\right)\left(n-m\right)}h_{n-m}\hat{a}_{n}^{\dagger}\hat{a}_{m}\right)\right]
\end{equation}
and
\begin{equation}
\hat{U}_{\text{int}}=\E^{-\I\hat{H}_{\text{int}}T/2}.
\end{equation}
Since $\hat{U}_{\text{int}}$ and $\hat{U}_{\text{nonint}}$\textbf{
}do not commute, the system will generally heat up to a featureless
stationary state. We note that the noninteracting part of the propagator
$\hat{U}_{\text{nonint}}$ corresponds to the propagator of \emph{a
noninteracting} system, exhibiting dynamical localization for appropriately
selected ratios of the driving amplitude (of the external potential)
to the driving frequency $\omega$. For example, for the hopping matrix,
\begin{equation}
h_{n-m}=\left(\delta_{n,m+1}+\delta_{n,m-1}\right)/2\label{eq:hopping-matrix}
\end{equation}
 and drive
\begin{equation}
f\left(t\right)=\begin{cases}
-A & -\frac{T}{2}\leq t<0\quad\left(\mod\,T\right)\\
A & 0\leq t<\frac{T}{2}\quad\left(\mod\,T\right),
\end{cases}\label{eq:driveterm}
\end{equation}
this condition is satisfied for \cite{Dunlap1988}
\begin{equation}
A/\omega=2n\qquad n\in\mathbb{Z\setminus}\{0\}.\label{eq:dynloc_pnt}
\end{equation}
In this case $\hat{U}_{\text{nonint}}$ becomes trivial, $\hat{U}_{\text{nonint}}=I$
\cite{Revie1986,Dunlap1988}, and the full propagator reduces to
\begin{equation}
\hat{U}\left(T,0\right)=\E^{-\I\hat{H}_{\text{int}}T/2}\label{eq:time_propagator}
\end{equation}
such that the stroboscopic dynamics evolves according to a simple
effective Hamiltonian which is identical to $\hat{H}_{\text{int}}$.
Since we require $\left[\hat{H}_{\text{int}},\hat{n}_{m}\right]=0$,
this propagator is diagonal in the position basis, implying a complete
dynamical localization of the model. One can easily show that for
interactions with finite support the spreading of any local operator
is finite and that for such models even entanglement does not grow
with time. The stroboscopic evolution of a many-body wavefunction
is generated by the repeated application of the one-period propagator
to the wavefunction
\begin{equation}
\ket{\psi\left(nT\right)}=\hat{U}^{n}\ket{\psi\left(0\right)}.\label{eq:psi_evol}
\end{equation}
 This evolution is generally quasiperiodic with multiple incommensurate
frequencies leading to correlation functions which generically decay
in time. However, periodic stroboscopic revivals of the many-body
wavefunction can be obtained if $\hat{U}$ satisfies $\hat{U}^{p}=I$
, such that its spectrum is concentrated on the complex $p-$roots
of unity (see Fig.~\ref{fig:spec}). While normally one has no direct
access to the properties of $\hat{U}$, for the systems we consider
here, its properties are directly determined by the eigenvalues of
$\hat{H}_{\text{int}}$ (see Eq.~(\ref{eq:time_propagator})). It
is easy to see that for $\hat{H}_{\text{int}}$ with eigenvalues,
$E_{\alpha}=J_{z}m_{\alpha}$ and $m_{\alpha}\in\mathbb{Z}$, the
propagator in (\ref{eq:time_propagator}) continued\emph{ }to\emph{
continuous time} is periodic with a period of $T'=4\pi/J_{z}$, that
is, $\hat{U}\left(t+T'\right)=\hat{U}\left(t\right)$.

If $T$ and $T'$ are commensurate, namely, $T/T'=q/p$ with $q,p\in\mathbb{Z}$
the stroboscopic dynamics is periodic with a subharmonic period of
$pT$, and the wavefunction \emph{exactly} revives after $p$ periods
of the drive. We will refer to this subharmonic revival mode as $(q:p)$.
In case that $T$ and $T'$ are \emph{incommensurate}, the revivals
are quasiperiodic, with the wavefunction coming back arbitrary close
to its initial state after sufficiently long times. We stress that
while in this case the eigenvalues of $\hat{U}$ ergodicaly cover
the unit circle (see Fig.~\ref{fig:spec}), since only two incommensurate
\emph{base} frequencies are involved, correlations function will \emph{not}
decay with time. 

(Quasi)periodic revivals of an arbitrary initial state mean that the
state does not obey the discrete time translational symmetry as the
Hamiltonian. It is however important to note that this symmetry is
not \emph{spontaneously} broken due to the absence of \emph{spatial}
order in our system \cite{Watanabe2015}.

The results of this section are rigorous, however the proposed driving
protocol is fine-tuned. It is therefore pertinent to examine the stability
of the revivals, which we do in the next section.

\emph{Stability analysis.\textemdash{}} To examine the stability of
the construction above, we numerically study two simple perturbations:
open boundary conditions and a static disordered potential. 

We set the hopping matrix $h_{n-m}$ in Eq.~(\ref{eq:ham}) as in
Eq.~(\ref{eq:hopping-matrix}) , the driving protocol as in (\ref{eq:driveterm})
and tune to the dynamical localization point (\ref{eq:dynloc_pnt})
with $A=4\pi/T$. The period of the drive is taken to be either $T=3$
or $4$ , such that the corresponding frequencies are much smaller
than the single-particle bandwidth. We also use a nearest neighbor
interaction of the form
\begin{equation}
\hat{H}_{\text{int}}=J_{z}\sum_{i}\hat{n}_{i}\hat{n}_{i+1}.
\end{equation}
As explained in the previous section, this system exhibits (quasi)periodic
oscillations of observables for (in)commensurate $T/\left(T'\right)=q/p$
with $T'=4\pi/J_{z}$. To examine the stability of a subharmonic revival
mode $\left(q:p\right)$, we calculate the discrete Fourier transform
of the one-particle Green's function computed at infinite temperature,
\begin{equation}
\tilde{G}_{i}(\omega)=\frac{1}{\sqrt{N}}\sum_{n=0}^{N}\E^{\I\omega Tn}\frac{1}{Z}\tr\left(\hat{c}_{i}^{\dagger}(nT)\hat{c}_{i}\right),\label{eq:greens_function}
\end{equation}
where $N$ is the number of periods we propagate. 

\begin{figure}
\includegraphics{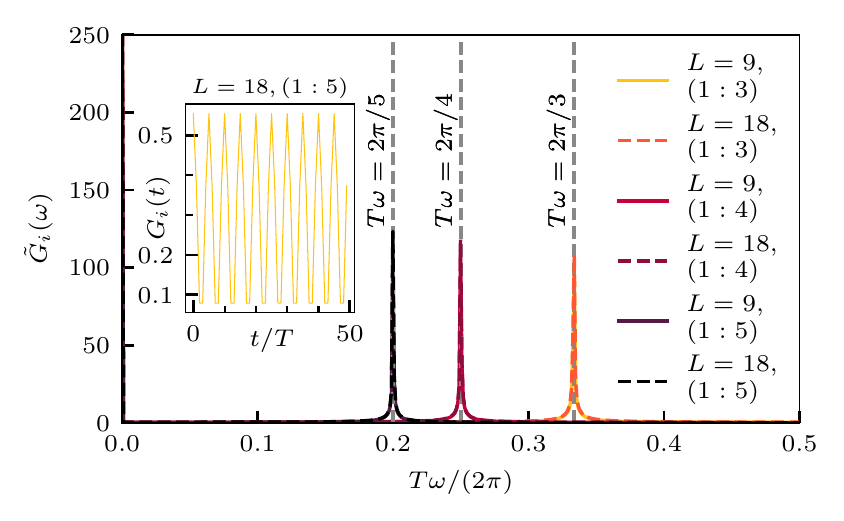}

\caption{Fourier transform of the one particle Green's function, $\tilde{G}_{i}\left(\omega\right)={\cal FT}\left[G_{i}(t)\right]$,
in a clean open chain of lengths $L=18$ and $L=9$ for different
interaction strengths tuned to different subharmonic modes $(1:p)$,
using a driving period of $T=4$. The inset shows the stroboscopic
dynamics in the real time domain for $p=5$, which is not periodic
in $T$, but periodic in $5T$\label{fig:boundary}}
\end{figure}

Instead of calculating the trace over the full Hilbert space, which
would require the full diagonalization of the Hamiltonian, we approximate
it by an expectation value calculated with respect to a random state,
$\ket{\psi_{0}}$, sampled randomly from the Haar measure \cite{Popescu2006}.
This approximation is of exponential precision in the dimension of
the Hilbert space \cite{Levy1939}. To further reduce the error we
average our results over 10-100 such random initial states. The calculation
of the one-particle Green's function is then reduced to the propagation
of two vectors $\ket{\psi_{L}\left(n\right)}=\hat{U}^{n}\ket{\psi_{0}}$
and $\ket{\psi_{R}\left(n\right)}=\hat{U}^{n}\hat{c}_{i}\ket{\psi_{0}}$,
according to the driving protocol (\ref{eq:driveterm}) illustrated
in Fig. \ref{fig:Driving-protocol}, and taking the expectation value
$G_{i}\left(n\right)=\left\langle \psi_{L}\left(n\right)|\hat{c}_{i}^{\dagger}|\psi_{R}\left(n\right)\right\rangle $.
We note that this requires to consider two sectors of the Hamiltonian,
effectively doubling the dimension of the Hilbert space. The repeated
application of $\hat{U}\ket{\psi_{0}}$ can be efficiently computed
utilizing the sparse structure of the problem \cite{Al-Mohy2011}
(for additional details on the method see Sec. VA in Ref.~\cite{Luitz2016c}).
This allows us to study systems of sizes up to $L=21$ for very late
times $t\geq1000$. 

Since the theoretical treatment of the previous section applies to
either periodic boundary conditions or infinite system sizes, it is
important to verify that the effect is robust for open boundary conditions.
In Fig. \ref{fig:boundary}, we show $\tilde{G}_{i}(\omega)$ for
a system with open boundary conditions and an interaction strength
tuned to several subharmonic revival modes, $\left(1:p\right)$. We
observe a perfect revival of the Green's function, leading to a sharp
peak in the Fourier transform exactly at $\omega T=1/p$, without
a significant dependence on the size of the system up to the accessible
number of driving periods. We have checked that this result prevails
for various driving frequencies $\omega$ smaller than the single
particle bandwidth.

\begin{figure}
\includegraphics{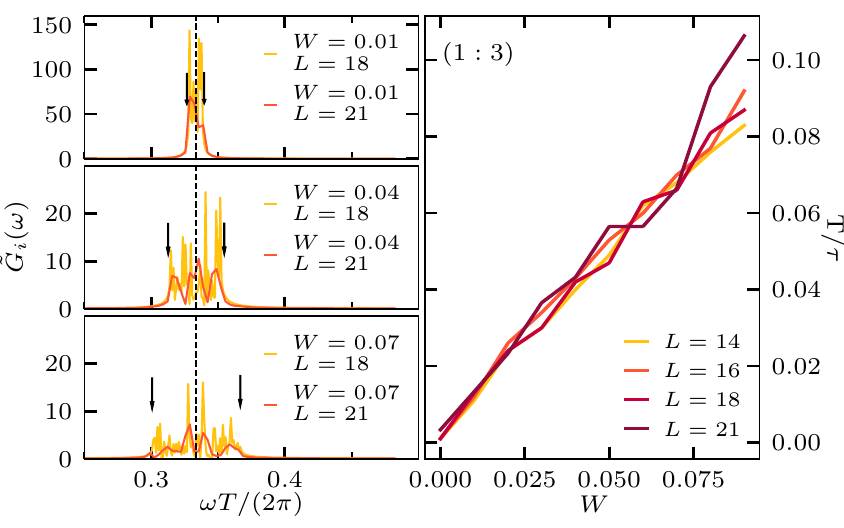}

\caption{\label{fig:stability_disorder}Stability of the subharmonic revival
mode $(1:3)$ to the introduction of static disorder. \textbf{Left:}
Fourier transform of the Green's function, $\tilde{G}_{i}\left(\omega\right)$
for different disorder strengths, $W$, for the subharmonic revival
mode $(1:3)$. Here, $T=3$ and $L=18$ ($L=21)$ using $1000$ ($300$)
driving periods respectively. Black arrows indicate the estimate of
the width of the revival mode for the $L=18$ data (see text) \textbf{Right:
}The width of the spectral peak on the left as a function of the disorder
strength. The visible small fluctuations originate both from statistical
and systematic sampling errors. }
\end{figure}

Any experimental realization of our system will be subject to imperfections.
To model such imperfections we introduce a static disorder potential
$\hat{H}_{\text{dis}}=\sum_{i}v_{i}\hat{n}_{i}$ with $\hat{v}_{i}$
uniformly distributed in the interval $\left[-W,W\right]$. In the
presence of disorder, the subharmonic revival modes acquire a life-time,
$\tau\left(W\right)$ which we estimate by measuring the broadening
of the corresponding spectral peaks. The width of the spectral peak
is taken as the maximal distance of satellite peaks (indicated by
black arrows in Fig. \ref{fig:stability_disorder}) which have a spectral
intensity of at least 10\% of the one of the central spectral peak.
We observe a linear dependence of the spectral width on the disorder
strength, which appears converged with the size of the system. The
life-time of the subharmonic revival modes therefore scales as $\tau\left(W\right)\sim1/W$.

\emph{Discussion.\textemdash }We have constructed a family of interacting
periodically driven systems exhibiting dynamical localization, which
can be tuned to have subharmonic periodic or quasiperiodic revivals
of the many-body wavefunction. Namely, all physical observables show
(quasi)periodic dependence, over time-scales \emph{longer} than the
period of the drive. The revivals are independent of the initial state
of the system and originate from the special structure of the eigenvalues
of the one-period propagator (see Fig.~\ref{fig:spec}). 

Despite not satisfying the conditions for spontaneous symmetry breaking
outlined in Ref.~\cite{Watanabe2015}, our construction serves as
a unique example of a nontrivial interacting quantum system with tunable
revivals. Our stability analysis shows that while the revivals are
stable under a change of boundary conditions, introduction of static
disorder leads to a finite lifetime inversely proportional to the
disorder strength. It is therefore possible that with proper control
of the disorder, subharmonic revivals could also be seen in cold atoms
experiments. 
\begin{acknowledgments}
YB acknowledges funding from the Simons Foundation (\#454951, David
R. Reichman). This project has received funding from the European
Union's Horizon 2020 research and innovation programme under the Marie
Sk\l odowska-Curie grant agreement No. 747914 (QMBDyn). DJL acknowledges
PRACE for awarding access to HLRS's Hazel Hen computer based in Stuttgart,
Germany under grant number 2016153659.
\end{acknowledgments}

\bibliographystyle{apsrev4-1}
\bibliography{subharmonic,lib_yevgeny,lib_tmp_acl}

\end{document}